\documentclass[runningheads]{llncs}
\usepackage[T1]{fontenc}
\usepackage{graphicx}
\usepackage{algorithm,algorithmic}
\usepackage{booktabs} 

\usepackage{pifont}
\newcommand{\cmark}{\ding{51}}
\newcommand{\xmark}{\ding{55}}

\begin{document}
\title{Segmentation of Kidney Tumors on Non-Contrast CT Images using Protuberance Detection Network}
%


\titlerunning{Segmentation of Kidney Tumors on NCCT Images}
%
\author{
    Taro Hatsutani \and
    Akimichi Ichinose\and
    Keigo Nakamura\and
    Yoshiro Kitamura
}

%
\authorrunning{T. Hatsutani et al.}

%
\institute{
    Medical Systems Research \& Development Center, FUJIFILM Corporation, Japan
    \email{taro.hatsutani@fujifilm.com}
}

%
\maketitle              

\begin{abstract}
    Many renal cancers are incidentally found on non-contrast CT (NCCT) images. On contrast-enhanced CT (CECT) images, most kidney tumors, especially renal cancers, have different intensity values compared to normal tissues. However, on NCCT images, some tumors called isodensity tumors, have similar intensity values to the surrounding normal tissues, and can only be detected through a change in organ shape. Several deep learning methods which segment kidney tumors from CECT images have been proposed and showed promising results. However, these methods fail to capture such changes in organ shape on NCCT images. In this paper, we present a novel framework, which can explicitly capture protruded regions in kidneys to enable a better segmentation of kidney tumors. We created a synthetic mask dataset that simulates a protuberance, and trained a segmentation network to separate the protruded regions from the normal kidney regions. To achieve the segmentation of whole tumors, our framework consists of three networks. The first network is a conventional semantic segmentation network which extracts a kidney region mask and an initial tumor region mask. The second network, which we name protuberance detection network, identifies the protruded regions from the kidney region mask. Given the initial tumor region mask and the protruded region mask, the last network fuses them and predicts the final kidney tumor mask accurately. The proposed method was evaluated on a publicly available KiTS19 dataset, which contains 108 NCCT images, and showed that our method achieved a higher dice score of 0.615 (+0.097) and sensitivity of 0.721 (+0.103) compared to 3D-UNet. To the best of our knowledge, this is the first deep learning method that is specifically designed for kidney tumor segmentation on NCCT images.

\end{abstract}

\section{Introduction}
Over 430,000 new cases of renal cancer were reported in 2020 in the world \cite{kidney_caner_number} and this number is expected to rise \cite{rcc_increase1}. When the tumor size is large (greater than 7cm) often the whole kidney is removed, however, when the tumor size is small (less than 4cm), partial nephrectomy is the preferred treatment \cite{PN_evidence} as it could preserve kidney's function. Thus, early detection of kidney tumors can help to improve patient's prognosis. However, early-stage renal cancers are usually asymptomatic, therefore they are often incidentally found during other examinations \cite{rcc_incidental1}, which includes non-contrast CT (NCCT) scans.

\begin{figure}[tb]
    \label{fig:ncc}
    \includegraphics[width=\linewidth]{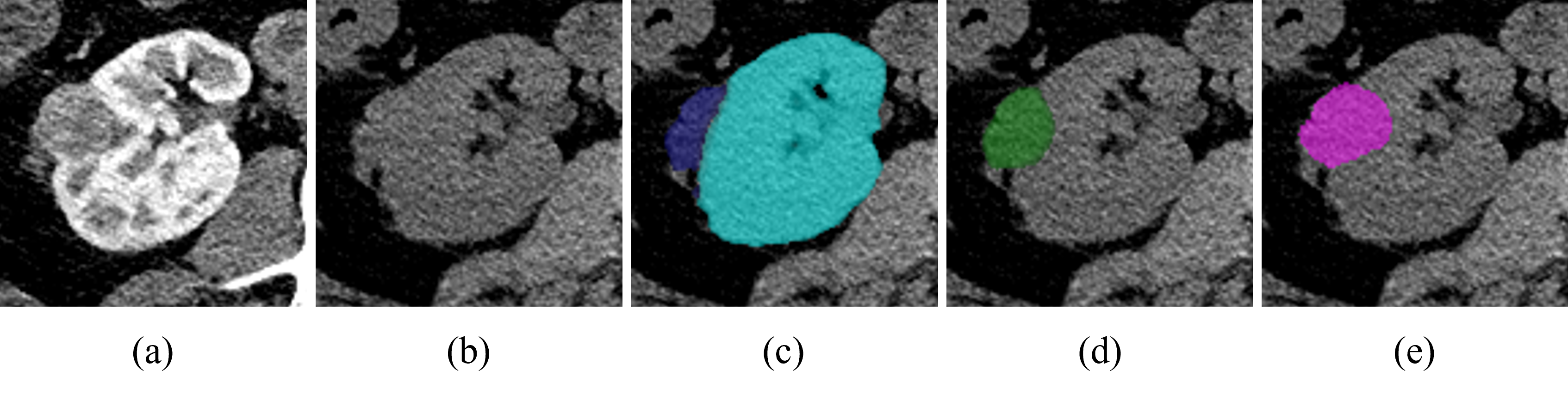}
    \caption{Example CECT and NCCT images. Some kidney tumors have similar intensity values to its surrounding tissues. a) CECT image. b) NCCT image. c) Output of the protuberance detection network. d) Output of our model. e) Ground truth mask.}
\end{figure}

Segmentation of kidney tumors on NCCT images adds challenges compared to contrast-enhanced CT (CECT) images, due to low contrast and lack of multi-phase images. On CECT images, the kidney tumors have different intensity values compared to the normal tissues. There are several works that demonstrated successful segmentation of kidney tumors with high precision \cite{kits19_1st,kits21_1st}. However, on NCCT images, as shown in Fig. \ref{fig:ncc}b, some tumors called isodensity tumors, have similar intensity values to the surrounding normal tissues. To detect such tumors, one must compare the kidney shape with tumors to the kidney shape without the tumors so that one can recognize regions with protuberance.

3D U-Net \cite{3D-UNet} is the go-to network for segmenting kidney tumors on CECT images. However, convolutional neural networks (CNNs) are biased towards texture features \cite{cnn_biased_texture}. Therefore, without any intervention, they may fail to capture the protuberance caused by isodensity tumors on NCCT images.

In this work, we present a novel framework that is capable of capturing the protuberances in the kidneys. Our goal is to segment kidney tumors including isodensity types on NCCT images. To achieve this goal, we create a synthetic dataset, which has separate annotations for normal kidneys and protruded regions, and train a segmentation network to separate the protruded regions from the normal kidney regions. In order to segment whole tumors, our framework consists of three networks. The first is a base network, which extracts kidneys and an initial tumor region masks. The second protuberance detection network receives the kidney region mask as its input and predicts a protruded region mask. The last fusion network receives the initial tumor mask and the protruded region mask to predict a final tumor mask. This proposed framework enables a better segmentation of isodensity tumors and boosts the performance of segmentation of kidney tumors on NCCT images. The contribution of this work is summarized as follows:

\begin{enumerate}
    \item Present a pioneering work for segmentation of kidney tumors on NCCT images.

    \item Propose a novel framework that explicitly captures protuberances in a kidney to enable a better segmentation of tumors including isodensity types on NCCT images. This framework can be extended to other organs (e.g. adrenal gland, liver, pancreas).

    \item Verify that the proposed framework achieves a higher dice score compared to the standard 3D U-Net using a publicly available dataset.
\end{enumerate}

\section{Related Work}
The release of two public CT image datasets with kidney and tumor masks from the 2019/2021 Kidney and Kidney Tumor Segmentation challenge \cite{kits19} (KiTS19, KiTS21) attracted researchers to develop various methods for segmentation.

Looking at the top 3 teams from each challenge \cite{kits19_1st,kits19_2nd,kits19_3rd,kits21_1st,kits21_2nd,kits19_3rd}, all teams utilized 3D U-Net \cite{3D-UNet} or V-Net \cite{V-Net}, which bears a similar architecture. The winner of KiTS19 \cite{kits19_1st} added residual blocks \cite{resnet} to 3D U-Net and predicted kidney and tumor regions directly. However, the paper notes that modifying the architecture resulted in only slight improvement. The other 5 teams took a similar approach to nnU-Net's coarse-to-fine cascaded network \cite{nnU-Net}, where it predicts from a low-resolution image in the first stage and then predicts kidneys and tumors from a high-resolution image in the second stage. Thus, although other attempts were made, using 3D U-Net is the go-to method for predicting kidneys and tumors. In our work, we also make use of 3D U-Net, but using this network alone fails to learn some isodensity tumors. To overcome this issue, we developed a framework that specifically incorporates protuberances in kidneys, allowing for an effective segmentation of tumors on NCCT images.

In terms of focusing on protruded regions in kidneys, our work is close to \cite{NCCT_manifold1,NCCT_manifold2}. \cite{NCCT_manifold2} developed a computer-aided diagnosis system to detect exophytic kidney tumors on NCCT images using belief propagation and manifold diffusion to search for protuberances. An exophytic tumor is located on the outer surface of the kidney that creates a protrusion. While this method demonstrated high sensitivity (95\%), its false positives per patient remained high (15 false positives per patient). In our work, we will not only segment protruded tumors but also other tumors as well.

\section{Proposed Method}
To capture the protuberances in kidneys, we specifically train a protuberance detection network, which receives a kidney region mask as an input and separates protruded regions from it. This enables us to extract a part of tumors that forms protuberance, but our goal is segmenting all visible kidney tumors on NCCT images. Thus, we make use of three networks as shown in Fig. 2.

The first base network is responsible for predicting kidney and tumor region masks. Our architecture is based on 3D U-Net, which has an encoder-decoder style architecture, with few modifications. To reduce the required size of GPU memory, we only use the encoder that has only 16 channels at the first resolution, but instead we make the architecture deeper by having 1 strided convolution and 4 max-pooling layers. In the decoder, we replace the up-convolution layers with a bilinear up-sampling layer and a convolution layer. In addition, by only having a single convolution layer instead of two in the original architecture at each resolution, we keep the decoder relatively small. Throughout this paper, we refer this architecture as our 3D U-Net.

The second protuberance detection network is the same as the base network except it starts from 8 channels instead of 16. We train this network using synthetic datasets. The details of the dataset and training procedures are described in section \ref{synthetic_dataset}.

The last fusion network combines the outputs from the base network and the protuberance detection network and makes the final tumor prediction. In detail, we perform a summation of the initial tumor mask and the protruded region mask, and then concatenate the result with the input image. This is the input of the last fusion network, which also has the same architecture as the base network with an exception of having two input channels. This fusion network do not just combine the outputs but also is responsible for removing false positives from the base network and the protuberance detection network.

Our combined three network is fully differentiable, however, to train efficiently, we train the model in 3 steps.

\begin{figure}[tbp]
    \label{fig:model}
    \includegraphics[width=\textwidth]{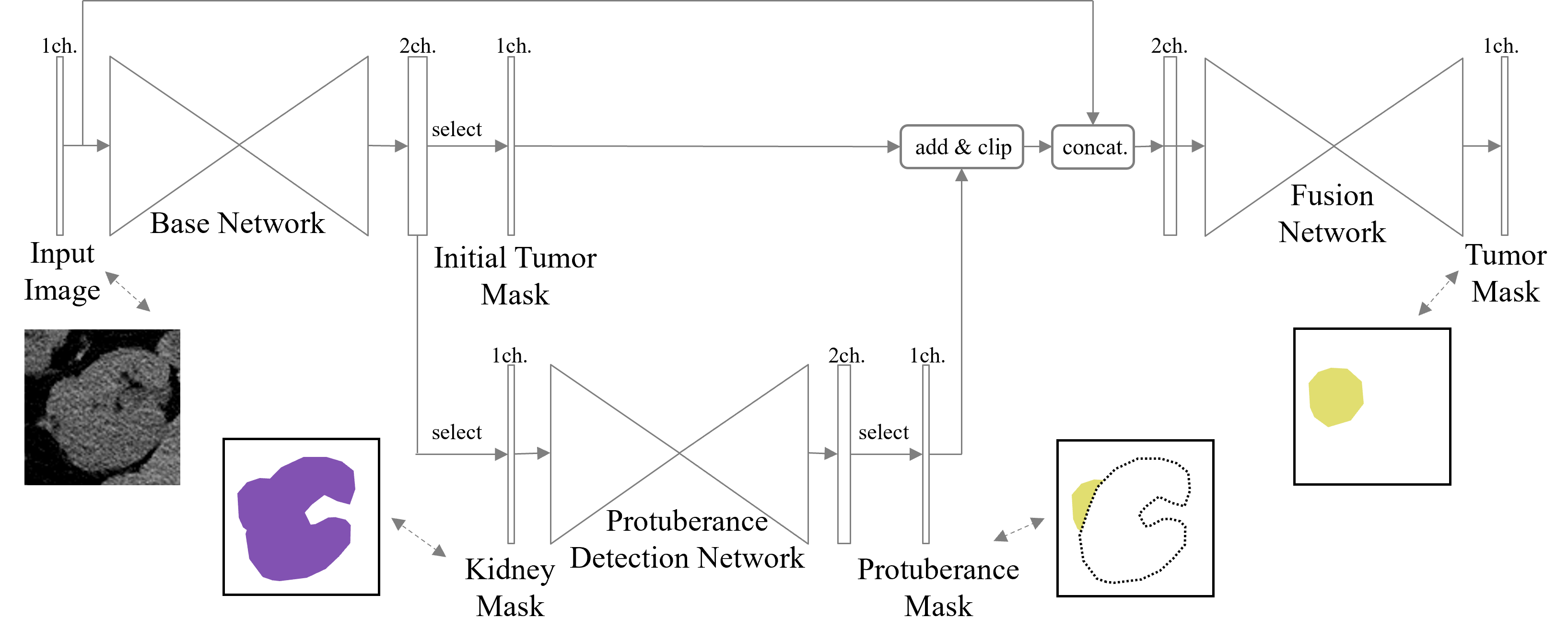}
    \caption{Overview of our framework.}
\end{figure}

\subsection{Step1: Training Base Network}
In the first step, we train the base network, which is a standard segmentation network, to extract kidney and tumor masks from the images. We use a sigmoid function for the last layer. And as a loss function, we use the dice loss \cite{V-Net} and the cross-entropy loss equally.

\subsection{Step2: Training Protuberance Detection Network}
In the second step, we train the protuberance detection network alone to separate protruded regions from the normal kidney masks. Here, we only use the cross-entropy loss and label smoothing with a smoothing factor of $\epsilon=0.01$. 

\subsubsection{Synthetic dataset} \label{synthetic_dataset}
To enable a segmentation of protruded regions only, a separate annotation of each region is usually required. However, annotating such areas is time-consuming and preparing a large number of data is challenging. Alternatively, we create a synthetic dataset that mimics a kidney with protrusions. The synthetic dataset is created through the following steps:
\begin{enumerate}
    \item Randomly sample a kidney mask without protuberance and a tumor mask.
    \item Apply random rotation and scaling to the tumor mask.
    \item Randomly insert the tumor mask into the kidney mask.
    \item If both of the following conditions are met, append to the dataset.
          \begin{equation}
              \label{eqn:cond1}
              \frac{\sum_i k_i t_i}{\sum_i k_i} < 0.3,
          \end{equation}
          \begin{equation}
              \label{eqn:cond2}
              \frac{\sum_i k_i t_i}{\sum_i t_i} < 0.95,
          \end{equation}
          where $k_i$ is a voxel value (0 or 1) in the kidney mask and $t_i$ is a voxel value in the tumor mask. Eq. \ref{eqn:cond1} ensures that only up to 30\% of the kidney is covered with a tumor. Eq. \ref{eqn:cond2} ensures that not all tumors are covered by the kidney (at least 5\% of the tumor is protruded from the kidney).
\end{enumerate}

\subsection{Step3: End-to-End Training with Fusion Network}
In the final step, we train the complete network jointly. Although our network is fully differentiable, since there is no separate annotation for protruded regions other from the synthetic dataset, we freeze the parameters in protuberance detection network.

The output of the protuberance detection network will likely have more false positives than the base network since it has no access to the input image. Thus, when the output of the protuberance detection network is concatenated with the output of the base network, the fusion network can easily reduce the loss by ignoring the protuberance detection network's output, which is suboptimal. To avoid this issue, we perform summation not concatenation to avoid the model from ignoring all output from the protuberance detection network. We then clip the value of the mask to the range of 0 and 1. As a result, the input to the fusion network has two channels. The first channel is the input image, and the second channel is the result of summation of the initial tumor mask and the protruded region mask. We concatenate the input image so that the last network can remove false positives from the predicted masks as well as predicting the missing tumor regions from the protuberance detection network.

We use the dice loss and the cross-entropy loss as loss functions for the fusion network. We also keep the loss functions in the base network for predicting kidneys and tumors. The loss function for tumors in the base network acts like an intermediate supervision. Our network shares some similarities with the stacked hourglass network \cite{hourglass} where the network consists of multiple U-Net like hourglass modules and has intermediate supervision at the end of each hourglass module. By having multiple modules in this manner, the network can fix the initial mistakes in early modules and corrects in later modules.

\section{Experiments}
No prior work exists that uses NCCT images from KiTS19 \cite{tcia_kits19,kits19}. Thus, we first created our baseline model and compared the performance with existing methods on CECT images. This allows us to ensure that our baseline model and training procedure is appropriate. We then trained the model using NCCT images and compared with our proposed method.

\subsection{Datasets and Preprocessing}
We used a dataset from KiTS19 \cite{kits19} which contains both CECT and NCCT images. For CECT images, there are 210 images for training and validation and, 90 images for testing. For NCCT images, there are 108 images, which are different series of the 210 images. The ground truth masks are only available for the 210 CECT images. Thus, we transfer the masks to NCCT images. This is achieved by extracting kidney masks and adjusting the height of each kidney. The ground truth mask contains a kidney label and a kidney tumor label. Cysts are not annotated separately and included in the kidney label on this dataset. The data can be downloaded from The Cancer Imaging Archive (TCIA) \cite{tcia,tcia_kits19}.

The images were first clipped to the intensity value range of [-90, 210] and normalized from -1 to 1. The voxel spacings were normalized to 1mm. During the training, the images were randomly cropped to a patch size of $128 \times 128 \times 128$ voxels. We applied random rotation, random scaling and random noise addition as data augmentation.

During the \textit{Step}2 phase of the training, where we used the synthetic dataset, we created 10,000 masks using the method from section \ref{synthetic_dataset}. We applied some augmentations during training to input masks to simulate the incoming inputs from the base network. The output of the base network is not binarized to keep gradient from flowing, so the values are in the range [0, 1] and the edge of kidneys are usually smooth. Therefore, we applied gaussian blurring, gaussian noise addition and intensity value shifting.

\subsection{Training details and evaluation metrics}
Our model was trained using SGD with a 0.9 momentum and a weight decay of 1e-7. We employed a learning rate scheduler, which we warm-up linearly from 0.0001 to 0.1 during the first 30\% (for \textit{Step}1 and \textit{Step}3) or 10\% (for \textit{Step}2) of the total training steps and decreased following the cosine decay learning rate. A mini-batch size of 8, 16 and 4 were used, and trained for 250k, 100k and 100k steps during \textit{Step}1 to 3 respectively. We conducted our experiments using JAX (v.0.4.1) \cite{jax} and Haiku (v.0.0.9) \cite{haiku}. We trained the model using a single NVIDIA RTX A5000 GPU.

For the experiment on CECT images, we used the dice score as our evaluation metrics following the same formula from KiTS19. For the experiment on NCCT images, we also evaluated the sensitivity and false positives per image (FPs/image). We calculated as true positive when the predicted mask has the dice score greater than 0.5, otherwise we calculated as false negative. On the other hand, false positives were counted when the predicted mask did not overlap with any ground truth masks.

\section{Results}
\begin{table}[tbp]
    \begin{center}
        \caption{Dice performance of existing method and our baseline model.
            Evaluated using CECT images from KiTS19. Composite dice is an average dice between kidney and tumor dice. The results were obtained by submitting our predicted masks to the Grand Challenge page.}
        \label{tab:cect}
        \begin{tabular}{l c c c}
            \toprule
            \textbf{Method}                          & \textbf{Composite Dice} & \textbf{Kidney Dice} & \textbf{Tumor Dice} \\
            \midrule
            Isensee and Maier-Hein \cite{kits19_1st} & 0.9123                  & 0.9737               & 0.8509              \\
            Our baseline model                       & 0.8832                  & 0.9728               & 0.7935              \\
            \bottomrule
        \end{tabular}
    \end{center}
\end{table}

\subsection{Performance on CECT Images}
To show that our model is properly tuned, we compare our baseline model with an existing method using CECT images. As can be seen from Table \ref{tab:cect}, our model showed comparable scores to the winner of KiTS19 challenge. We used this baseline model as our base network for the experiments on NCCT images.

\subsection{Performance on NCCT Images}
\begin{figure}[tb]
    \centering
    \includegraphics[width=0.75\linewidth]{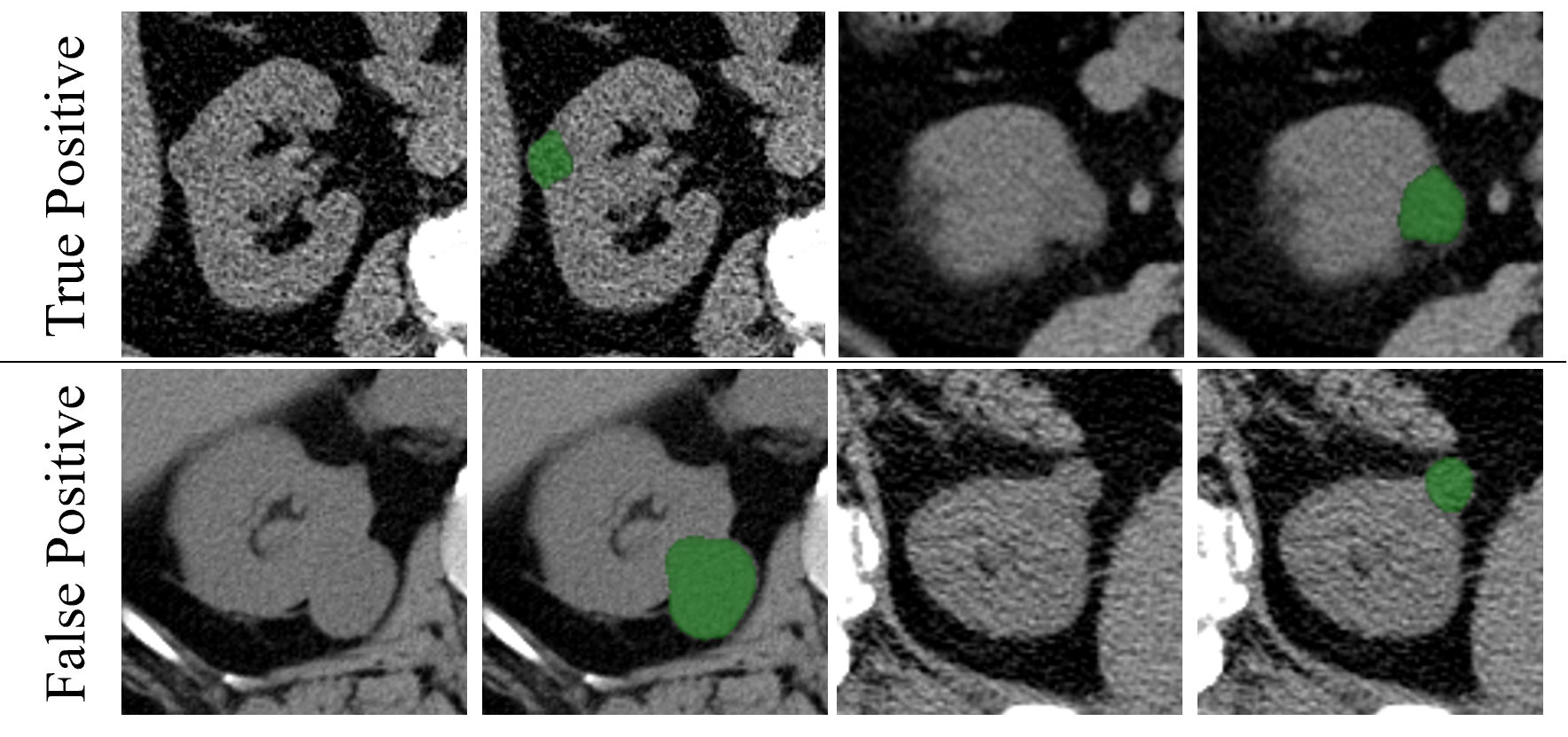}
    \caption{Output examples from our model. The first row shows true positive results and the second row shows false positive results.}
    \label{fig:tp_fp}
\end{figure}

Table \ref{tab:ncct} shows our experimental results and ablation studies on NCCT images. The proposed method (Table \ref{tab:ncct}-bottom) outperformed the baseline model (Table \ref{tab:ncct}-top). The ablation studies show that adding each component (CECT images and the protuberance detection network) resulted in an increase in the performance. While adding CECT images contributed the most for the increase in tumor dice and sensitivity, adding the protuberance detection network further pushed the performance. However, the false positives per image (FPs/image) increased from 0.283 to 0.421. The protuberance detection network cannot distinguish the protrusions that were caused by tumors or cysts, so the output from this network has many FPs at this stage. Thus, the fusion network has to eliminate cysts by looking again the input image, however, it may have failed to eliminate some cysts (Fig. \ref{fig:tp_fp} second row).

\begin{table}[tbp]
    \begin{center}
        \caption{Result of our proposed method on NCCT images from KiTS19. The values  are average values of a five-fold cross-validation.}
        \label{tab:ncct}
        \begin{tabular}{c c c c c c}
            \toprule
            \textbf{Protuberance}      & \textbf{with CECT} & \textbf{Tumor Dice} & \textbf{Sensitivity} & \textbf{FPs/Image} \\
            \textbf{Detection Network} & \textbf{images}    & \textbf{}           & \textbf{}                                 \\
            \midrule
            \xmark                     & \xmark             & 0.518               & 0.618                & \textbf{0.283}     \\
            \xmark                     & \cmark             & 0.585               & 0.686                & 0.340              \\
            \cmark                     & \cmark             & \textbf{0.615}      & \textbf{0.721}       & 0.421              \\
            \bottomrule
        \end{tabular}
    \end{center}
\end{table}

\section{Conclusion}
In this paper, we proposed a novel framework for kidney tumor segmentation on NCCT images. To cope with isodensity tumors, which have similar intensity values to their surrounding tissues, we created a synthetic dataset to train a network that extracts protuberance from the kidney masks. We combined this network with the base network and fusion network. We evaluated our method using the publicly available KiTS19 dataset, and showed that the proposed method can achieve a higher sensitivity than existing approach. Our framework is not limited to kidney tumors but can also be extended to other organs (e.g., adrenal gland, liver, pancreas).

\bibliographystyle{splncs04}
\bibliography{paper2085}

\begin{thebibliography}{10}
\providecommand{\url}[1]{\texttt{#1}}
\providecommand{\urlprefix}{URL }
\providecommand{\doi}[1]{https://doi.org/#1}

\bibitem{kidney_caner_number}
Estimated number of new cases in 2020, world, both sexes, all ages (excl.
  nmsc), \url{https://gco.iarc.fr/today/online-analysis-table/}, last accessed
  27 Feb 2023

\bibitem{jax}
Bradbury, J., Frostig, R., Hawkins, P., Johnson, M.J., Leary, C., Maclaurin,
  D., Necula, G., Paszke, A., Vander{P}las, J., Wanderman-{M}ilne, S., Zhang,
  Q.: {JAX}: composable transformations of {P}ython+{N}um{P}y programs (2018),
  \url{http://github.com/google/jax}

\bibitem{3D-UNet}
{\c{C}}i{\c{c}}ek, {\"O}., Abdulkadir, A., Lienkamp, S.S., Brox, T.,
  Ronneberger, O.: 3d u-net: Learning dense volumetric segmentation from sparse
  annotation. In: International Confernce on Medical Image Computing and
  Computer-Assisted Intervention. pp. 424--432. Springer (2016)

\bibitem{tcia}
Clark, K., Vendt, B., Smith, K., Freymann, J., Kirby, J., Koppel, P., Moore,
  S., Phillips, S., Maffitt, D., Pringle, M., Tarbox, L., Prior, F.: The cancer
  imaging archive (tcia): Maintaining and operating a public information
  repository. Journal of Digital Imaging  \textbf{26}(6),  1045--1057 (2013)

\bibitem{cnn_biased_texture}
Geirhos, R., Rubisch, P., Michaelis, C., Bethge, M., Wichmann, F.A., Brendel,
  W.: Imagenet-trained {CNN}s are biased towards texture; increasing shape bias
  improves accuracy and robustness. In: Proceedings of International Conference
  on Learning Representations (2019)

\bibitem{kits21_2nd}
Golts, A., Khapun, D., Shats, D., Shoshan, Y., Gilboa-Solomon, F.: An ensemble
  of 3d u-net based models for segmentation of kidney and masses in ct scans.
  In: Kidney and Kidney Tumor Segmentation. pp. 103--115. Springer (2022)

\bibitem{resnet}
He, K., Zhang, X., Ren, S., Sun, J.: Deep residual learning for image
  recognition. In: Proceedings of the IEEE Conference on Computer Vision and
  Pattern Recognition. pp. 770--778 (2016)

\bibitem{kits19}
Heller, N., Isensee, F., Maier-Hein, K.H., Hou, X., Xie, C., Li, F., Nan, Y.,
  Mu, G., Lin, Z., Han, M., Yao, G., Gao, Y., Zhang, Y., Wang, Y., Hou, F.,
  Yang, J., Xiong, G., Tian, J., Zhong, C., Ma, J., Rickman, J., Dean, J.,
  Stai, B., Tejpaul, R., Oestreich, M., Blake, P., Kaluzniak, H., Raza, S.,
  Rosenberg, J., Moore, K., Walczak, E., Rengel, Z., Edgerton, Z., Vasdev, R.,
  Peterson, M., McSweeney, S., Peterson, S., Kalapara, A., Sathianathen, N.,
  Papanikolopoulos, N., Weight, C.: The state of the art in kidney and kidney
  tumor segmentation in contrast-enhanced ct imaging: Results of the kits19
  challenge. Medical Image Analysis  \textbf{67},  101821 (2021)

\bibitem{tcia_kits19}
Heller, N., Sathianathen, N., Kalapara, A., Walczak, E., Moore, K., Kaluzniak,
  H., Rosenberg, J., Blake, P., Rengel, Z., Oestreich, M., Dean, J., Tradewell,
  M., Shah, A., Tejpaul, R., Edgerton, Z., Peterson, M., Raza, S., Regmi, S.,
  Papanikolopoulos, N., Weight, C.: C4kc kits challenge kidney tumor
  segmentation dataset (2019)

\bibitem{haiku}
Hennigan, T., Cai, T., Norman, T., Babuschkin, I.: {H}aiku: {S}onnet for {JAX}
  (2020), \url{http://github.com/deepmind/dm-haiku}

\bibitem{kits19_2nd}
Hou, X., Chunmei, X., Li, F., Yang, N.: Cascaded semantic segmentation for
  kidney and tumor. Submissions to the 2019 Kidney Tumor Segmentation
  Challenge: KiTS19  (2019)

\bibitem{nnU-Net}
Isensee, F., Jaeger, P.F., Kohl, S.A.A., Petersen, J., Maier-Hein, K.H.:
  {nnU}-net: a self-configuring method for deep learning-based biomedical image
  segmentation. Nature Methods  \textbf{18}(2),  203--211 (2020)

\bibitem{kits19_1st}
Isensee, F., Maier-Hein, K.: An attempt at beating the 3d u-net. Submissions to
  the 2019 Kidney Tumor Segmentation Challenge: KiTS19  (2019)

\bibitem{NCCT_manifold2}
Liu, J., Wang, S., Linguraru, M.G., Yao, J., Summers, R.M.: Computer-aided
  detection of exophytic renal lesions on non-contrast ct images. Medical Image
  Analysis  \textbf{19}(1),  15--29 (2015)

\bibitem{NCCT_manifold1}
Liu, J., Wang, S., Yao, J., Linguraru, M.G., Summers, R.M.: Manifold diffusion
  for exophytic kidney lesion detection on non-contrast ct images. In:
  International Confernce on Medical Image Computing and Computer-Assisted
  Intervention. pp. 340--347. Springer (2013)

\bibitem{V-Net}
Milletari, F., Navab, N., Ahmadi, S.A.: V-net: Fully convolutional neural
  networks for volumetric medical image segmentation. In: 2016 Fourth
  International Conference on 3D Vision (3DV). pp. 565--571. IEEE (2016)

\bibitem{kits19_3rd}
Mu, G., Lin, Z., Han, M., Yao, G., Gao, Y.: Segmentation of kidney tumor by
  multi-resolution vb-nets. Submissions to the 2019 Kidney Tumor Segmentation
  Challenge: KiTS19  (2019)

\bibitem{hourglass}
Newell, A., Yang, K., Deng, J.: Stacked hourglass networks for human pose
  estimation. In: Proceedings of European Conference on Computer Vision. pp.
  483--499. Springer (2016)

\bibitem{rcc_incidental1}
Pinsky, P.F., Dunn, B., Gierada, D., Nath, P.H., Munden, R., Berland, L.,
  Kramer, B.S.: {Incidental renal tumours on low-dose CT lung cancer screening
  exams}. J Med Screen  \textbf{24}(2),  104--109 (2017)

\bibitem{PN_evidence}
Touijer, K., Jacqmin, D., Kavoussi, L.R., Montorsi, F., Patard, J.J., Rogers,
  C.G., Russo, P., Uzzo, R.G., {Van Poppel}, H.: The expanding role of partial
  nephrectomy: A critical analysis of indications, results, and complications.
  European Urology  \textbf{57}(2),  214--222 (2010)

\bibitem{kits21_1st}
Zhao, Z., Chen, H., Wang, L.: A coarse-to-fine framework for the 2021 kidney
  and kidney tumor segmentation challenge. In: Kidney and Kidney Tumor
  Segmentation. pp. 53--58. Springer (2022)

\bibitem{rcc_increase1}
Znaor, A., Lortet-Tieulent, J., Laversanne, M., Jemal, A., Bray, F.:
  International variations and trends in renal cell carcinoma incidence and
  mortality. European Urology  \textbf{67}(3),  519--530 (2015)

\end{thebibliography}

\end{document}